\documentclass[a4paper]{spie}  

\usepackage[]{graphicx}

\title{Push button generation of multiphoton entanglement}
\author{Yuan Liang Lim\supit{a} and Almut Beige\supit{b,a}  
\skiplinehalf
\supit{a}Blackett Laboratory, Imperial College London, Prince Consort Road, London SW7 2BW, UK \\
\supit{b}Centre for Quantum Computation, Department of Applied Mathematics and Theoretical Physics, University of Cambridge, Wilberforce Road, Cambridge CB3 0WA, UK}

\authorinfo{Further author information: (Send correspondence to Y.L.L.) \\ Y.L.L.: E-mail: yuan.lim@imperial.ac.uk \\  A.B.: E-mail: a.beige@imperial.ac.uk}

\begin{document}

\maketitle

\begin{abstract}
Photon entanglement is an essential ingredient for linear optics quantum computing schemes, quantum cryptographic protocols and fundamental tests of 
quantum mechanics. Here we describe a setup that allows for the generation of polarisation-entangled $N$-photon states on demand. The photons are obtained by mapping the entangled state of $N$ atoms, each of them trapped inside an optical cavity, onto the 
free radiation field. The required initial state can be prepared by performing postselective measurements on the collective emission from the cavities through a multiport beamsplitter.
\end{abstract}


\section{Introduction}

Entanglement generation, be it for atoms or photons, spurs a great deal of interest in quantum information processing\cite{Gottesman} and quantum 
cryptography\cite{Bennett,Ekert}. Fundamental tests of quantum mechanics often require the presence of highly entangled states. For many 
practical purposes, the polarisation states of photons provide the most favoured qubits\cite{Knill}. The reason is that photons possess very long 
lifetimes and an ease in distribution. The main disadvantage of employing photons is that it is difficult to create an effective interaction between them and 
hence they are difficult to entangle. Nevertheless, in recent years, much progress has been made in the development of sources for the generation of 
multiphoton entanglement\cite{Kwiat} and the generation of single photons on demand\cite{Kuhn2,Yamamoto,KimbleNature}. 

The most common way to create mulitphoton entanglement is to generate the photons within the same source. This applies to parametric downconversion\cite{Kwiat}, which is currently used in many experiments, and to the atomic cascade\cite{Aspect}, which was the first source demonstrating a violation of Bell's famous inequality\cite{Bell}. Parametric downconversion schemes are in principle scalable to multiphoton entanglement.  However, this is intrinsically inefficient\cite{Bouwmeester,Eibl}.

Another way to {\em probabilistically} produce multiphoton entanglement is to create single photons on demand using strongly coupled atom-cavity systems\cite{Kuhn2,Maunz,KimbleNature} or quantum dot technology\cite{Yamamoto}. Afterwards these photons are brought to overlap on a beamsplitter within their coherence time 
and a successful postselective and entangling measurement should be performed on the output ports\cite{Fattal}. Schemes for the generation of photon 
entanglement with linear-optics setups are widely known\cite{Linear,Linear2,Linear3} and offer the advantage of being able to tailor a variety of 
multiphoton states. The main disadvantage of these schemes is that they are non-deterministic and the probability for the successful generation of a 
certain multiphoton state decreases in general rapidly with the number of photons involved.  

Schemes that are less flexible but allow for the {\em deterministic} generation of multiphoton entanglement have been proposed by Gheri {\em et 
al.}\cite{Gheri} and Lange and Kimble\cite{Lange}. Both proposals require only a single atom-cavity system and employ certain features of the photon 
generation process. The basic idea is to prepare the atom in a superposition of ground states which is then mapped step by step onto the state of each 
newly generated photon. During this process, the state of the first photon is highly entangled with the state of the atom inside the resonator. The state of each subsequently created photon can thus become strongly correlated to the state of all the others. In order to produce $N$ entangled photons usuable as qubits, the schemes\cite{Gheri,Lange} require to subsequently perform at least $N$ coherent operations on the same atom-cavity system and they are therefore relatively sensitive to decoherence in the case of the generation of several photons. 

Here, we describe a scheme that allows for the generation of $N$ highly entangled photons in $N$ spatially 
separated modes of the free radiation field in a {\em push button}-like experiment. We are able to engineer a wide range of multiphoton states which could then directly be used as an input for linear optics quantum computing schemes or quantum cryptography. The experimental setup consists of $N$ identical atom-cavity systems\footnote{How strictly this condition has to be fulfilled will be discussed elsewhere\cite{Lim}.} and we assume that the frequency is for all photons the same. The proposed scheme is very robust and requires essentially only the following two steps. Initially, the atoms have to be prepared in a highly entangled stable ground state.  This initialisation step can be realised with the help of postselective measurements on the collective emission from the cavities through a multiport beamsplitter. Once successfull, another set of $N$ photons can be generated using the same procedure as for the generation of single photons on demand\cite{Law,Kuhn}. During this step, the state of the atoms is mapped onto the state of $N$ newly generated photons. 

As in the entangled photon pair creation scheme by Lim and Beige\cite{LimPRL}, there is no need for the photons to be emitted at the same time. 
Furthermore, the adiabatic passage used for the excitation of the cavity mode can be performed with a very high fidelity\cite{Kuhn2}. The scheme is 
resilient against phase errors, i.e.~there is no need to control the relative phases between the different atom-cavity systems, since the initial state is a product state and phase fluctuations only affect a global phase with no physical consequences.

\section{Single photons on demand} \label{demand}

\begin{figure}
\begin{center}
\begin{tabular}{c}
\includegraphics[height=2.4cm]{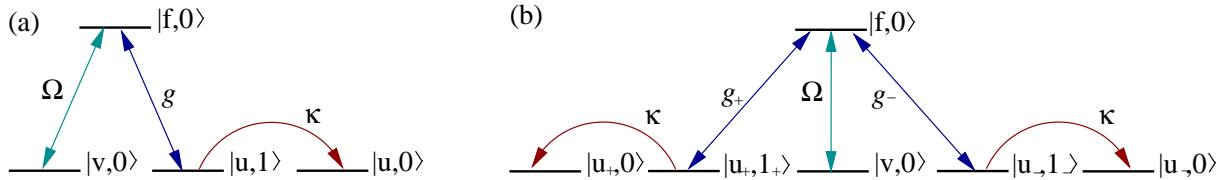}
\end{tabular}
\end{center}
\caption[example] { \label{stirap}  
(a) Level configuration for the generation of a single photon on demand using a stimulated Raman adiabatic passage (STIRAP)\cite{Bergmann}. This process 
consists of a counterintuitive pulse sequence that transfers $|v,0 \rangle$ adiabatically into the state $|u,1\rangle$. Afterwards, a photon leaks out 
through the cavity mirrors into the reservoir mode. (b) If the atom couples to the two different polarisation modes of the cavity field simultaneously, 
the counterintuitive pulse sequence transfers the system into a superposition of the states $|u_+,1_+\rangle$ and $|u_-,1_-\rangle$.}
\end{figure}  

Let us first discuss in detail one of the main ingredients for the {\em push button}-like generation of multiphoton entanglement, namely a scheme for 
the generation of single photons on demand\cite{Law,Kuhn}. Creating one photon requires only a single atom 
trapped inside an optical cavity and has already been realised experimentally by Kuhn {\em et al.}\cite{Kuhn2} and, more recently, by McKeever {\em et al.}\cite{KimbleNature} The relevant level configuration of the atom is shown in Figure \ref{stirap}(a) and the atom should initially be 
prepared in the atomic ground state $|v \rangle$ with no photon in the resonator. In the following we denote this state as $|v,0 \rangle$. The success 
probability of the scheme can be arbitrarily close to unity\cite{Kuhn} if the system is operated in the strong coupling regime where 
\begin{equation} \label{strong}
g^2 / \kappa \Gamma \gg 1 ~. 
\end{equation}
Here $\kappa$ and $\Gamma$ are the spontaneous decay rates of a single photon in the resonator field and the excited atomic state $|f \rangle$, respectively, 
while $g$ describes the atom-cavity coupling.

In order to create only one photon, it is important that there is only one atom in the cavity at a time. The basic idea is to induce a transition from the 
state $|v,0 \rangle$ directly to $|u,1 \rangle$ without ever populating the excited state $|f,0 \rangle$. This can be realised using a stimulated Raman 
adiabatic passage (STIRAP)\cite{Bergmann} based on the application of a so-called counterintuitive pulse sequence. Initially, the atom should interact with 
the cavity mode while the Rabi frequency $\Omega$ of the laser is zero. Then $\Omega$ should increase slowly until $\Omega/g \gg 1$. 

That this process indeed places one photon into the cavity mode can be understood with the help of the adiabatic theorem\cite{Law,Kuhn}. It states that a 
system, which is initially prepared in an eigenstate of the governing Hamiltonian, follows this eigenstate adiabatically if the parameters of the system 
change slowly in time. The Hamiltonian of the system we consider here equals 
\begin{equation} 
H  =  {\textstyle {1 \over 2}} \hbar \Omega \, |v \rangle \langle f| + \hbar g \, a^\dagger \, |f \rangle \langle u| +  {\rm H.c.} ~,
\end{equation} 
where $a^\dagger$ is the creation operator for a single photon in the cavity field. If the system is initially prepared in the zero eigenstate $|v,0 
\rangle$, it  evolves into
\begin{equation}
|\phi \rangle =\cos \vartheta \, |v,0 \rangle - \sin \vartheta \, |u,1 \rangle 
\end{equation}
with  $\tan \vartheta = \Omega/g$. When $\Omega/g \gg 1$, the relevant eigenstate of the system becomes $|u,1 \rangle$ and the time evolution of the system 
can effectively be summarised by the operator
\begin{equation} \label{U}
P_{\rm STIRAP} = |u,1 \rangle  \langle v,0| ~.
\end{equation}
During the population transfer from one atomic ground state into another, the excited state $|f \rangle$ accumulates only a very small population. This 
population is the smaller the slower $\Omega$ increases in time. Spontaneous emission from the atom into unwanted spatial directions remains therefore 
negligible. Once in the cavity, the photon leaks out through the outcoupling  mirror of the resonator with the cavity decay rate $\kappa$. To reinitialise the system, another laser pulse sequence transfers the system back into $|v,0 \rangle$. 

\begin{figure}
\begin{center}
\begin{tabular}{c}
\includegraphics[height=3.2cm]{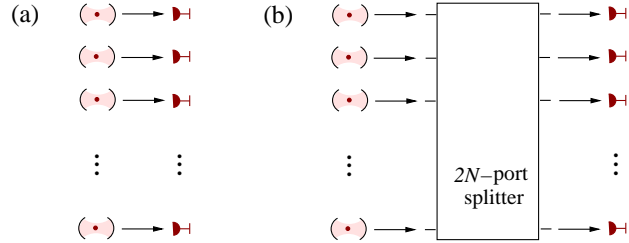}
\end{tabular}
\end{center}
\caption[example]{ \label{setup} 
Experimental setup. (a) During the final {\em push button} step, the entanglement of $N$ atom-cavity systems is mapped onto the state of $N$ newly 
generated photons. (b) The initialisation of the system requires postselective measurements on the photon emission from the $N$ cavities through a 
multiport beamsplitter.}
\end{figure} 

In the following we also consider a straightforward generalisation of this scheme where the atom simultaneously couples to the two different 
polarisation modes of the cavity field as shown in Figure \ref{stirap}(b). This allows us to determine the polarisation of the created photon by choosing 
the atom-cavity coupling constants $g_+$ and $g_-$ accordingly. Suppose the generation of a ``$+$'' polarised photon in the cavity corresponds to a transition from $|v,0 
\rangle$ to a state $|u_+,1 \rangle$ and  the generation of a ``$-$'' polarised photon to a transition from $|v,0 \rangle$ to a state $|u_-,1 \rangle$. Then the STIRAP operator becomes in analogy to Eq.~(\ref{U})
\begin{equation} \label{U2}
\tilde P_{\rm STIRAP} = {\textstyle {1 \over \sqrt{2}}} \, \big[ |u_+,1_+ \rangle + |u_-,1_- \rangle \big] \langle v,0| ~,
\end{equation}
if the atom-cavity coupling constants $g_+$ and $g_-$ are for both field modes the same. Here the state $|1_\pm \rangle$ denotes a one-photon state with 
one ``$+$'' or one ``$-$'' polarised photon, respectively.

\section{Experimental setup and basic idea}

In the following we consider STIRAP processes to explore the possiblity of generating $N$ spatially separated and polarisation-entangled photons in a {\em push button}-like experiment. The proposed setup requires a multiport beamsplitter and photon detectors  as well as $N$ distant optical cavities, each of them containing one atom (see Figure \ref{setup}). As in Section \ref{demand}, we assume the strong coupling regime (\ref{strong}). In addition, the cavity decay rate $\kappa$ should be for both polarisation modes ``$+$''  and ``$-$'' and for all cavities the same.  

Using this setup, the generation of multiphoton entanglement consists essentially of two steps. First, the atoms are prepared in a highly entangled state by performing postselective measurements on the 
collective emission from the cavities through the multiport beamsplitter (see Figure \ref{setup}(b)). This step is based on interference effects and postselective measurements as 
reported for example in Refs.\cite{LimPRL,Cabrillo,Cabrillo2,Cabrillo3,Cabrillo4,Cabrillo5} The final  {\em push-button} step consists of the simultaneous creation of $N$ photons (see Figure \ref{setup}(a)). Thereby the state of the atoms is mapped onto the state of the newly generated photons\cite{Parkins,Parkins2}. 

\begin{figure}
\begin{center}
\begin{tabular}{c}
\includegraphics[height=2.8cm]{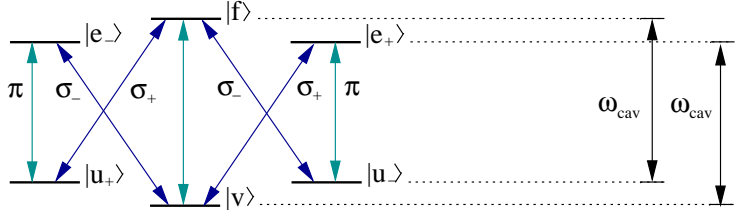}
\end{tabular}
\end{center}
\caption[example]{ \label{levelsun} 
Relevant level configuration of the atom trapped inside an optical cavity.}
\end{figure}

Being able to perform these two steps requires a relatively complex atomic level configuration (see Figure \ref{levelsun}). Suppose the atoms are initially all prepared in $|v,0 \rangle$. Then the initialisation step consists of a resonant $\pi$-polarised laser field which couples the atomic 
ground state $|v \rangle$ to the excited state  $|f \rangle$. During the mapping, a $\pi$-polarised laser drives the transitions between the levels $|u_+ 
\rangle$, $|u_- \rangle$ to the excited states $|e_- \rangle$, $|e_+ \rangle$, respectively. Once excited, the atom of each atom-cavity system couples via the $|f \rangle-|u_+ 
\rangle$ transition and the $|e_+ \rangle-|v \rangle$ transition to a state with one ``$+$'' polarised photon in the corresponding resonator. During a $|f \rangle-|u_- 
\rangle$ or a $|e_- \rangle-|v \rangle$ transition a ``$-$" polarised photon is created.

\subsection{Hamilton description of the atom-cavity system} \label{Ham}

If $\Omega^{(i)}_x$ denotes the Rabi frequency of the laser exciting the ground state $|x \rangle$ in atom $i$, then the laser Hamiltonian for atom-cavity 
subsystem $i$ becomes in the interaction picture with respect to the free Hamiltonian
\begin{equation} \label{laser}
H_{\rm laser}^{(i)}  = {\textstyle {1 \over 2}} \hbar \Omega^{(i)}_v \, |v \rangle_{ii} \langle f| +  {\textstyle {1 \over 2}} \hbar \Omega^{(i)}_{u_+} \, 
|u_+ \rangle_{ii} \langle e_-| + {\textstyle {1 \over 2}} \hbar \Omega^{(i)}_{u_-} \, |u_- \rangle_{ii} \langle e_+| + {\rm H.c.} 
\end{equation}
Note that the Rabi frequencies are in general time dependent. They are varied in time in order to realise adiabatic passages as described in Section 
\ref{demand}. In this paper,  we assume the energy levels $|u_+ \rangle$ and $|u_- \rangle$ to be degenerate. The same should apply to the states  $|e_+ \rangle$ 
and $|e_-\rangle$. In order to describe the atom-cavity interaction, we denote the creation operator for a single photon with polarisation $\lambda$ in cavity $i$ by 
$a_{\lambda}^{(i)\dagger}$ while $g^{(i)}_{x}$ is the atom-cavity coupling constant for a transition involving the atomic state $|x \rangle_i$. Then the 
atom-cavity interaction Hamiltonian for subsystem $i$ becomes in the interaction picture with respect to the free Hamiltonian 
\begin{equation} \label{cav}
H_{\rm cav}^{(i)}  = \sum_{\lambda=+,-}  \hbar g^{(i)}_{u_\lambda} \, a^{(i) \dagger}_{\lambda} \, |u_\lambda \rangle_{ii} \langle f| +  \hbar 
g^{(i)}_{e_\lambda} \, a_{\lambda}^{(i) \dagger} \, |v \rangle_{ii} \langle e_\lambda|  
+ {\rm H.c.} 
\end{equation} 
Here we assumed that each atom couples to two different polarisation modes of the resonator field.

\subsection{Leakage of photons through the cavity mirrors}

Suppose $b^{(i) \dagger}_\lambda$ denotes the creation operator for a single photon with polarisation $\lambda$ beyond the outcoupling mirror of cavity 
$i$.  Then the creation of a photon in this mode and the corresponding annihilation of a photon with the same polarisation inside resonator $i$ can be 
described by the leakage operator $L^{(i)}$ with
\begin{equation} \label{reset}
L^{(i)} = \sum_{\lambda=+,-} a^{(i)}_{\lambda} b_{\lambda}^{(i)\dagger} ~.
\end{equation}
Applied to the state of the system, this operator describes (up to normalisation) the effect of the leakage of a photon through the outcoupling mirror of cavity $i$ without affecting its polarisation. 

\subsection{Transition of photons through the linear optics setup}

In order to take the effect of the presence of the  multiport beamsplitter during the initialisation step into account we now introduce the leakage 
operator $\tilde L^{(i)}$ which describes the transition of a photon from cavity $i$ to the $N$ output ports of the multiport beamsplitter. To find an 
expression for this operator we use the scattering theory\cite{Pryde} and denote the probability amplitude with which a single photon is  redirected from 
input port $i$ to output port $j$ by $U_{ij}$. Let us denote the creation operator for a single photon with polarisation $\lambda$ in output port $j$ of 
the linear optics setup by $c_{\lambda}^{(j)\dagger}$ while $S$ is the scattering operator associated with the setup. Then the  input photon operator $b^{(i)}_\lambda$ in Eq.~(\ref{reset}) has to be replaced by $S \, b_{\lambda}^{(i)\dagger} \, S^\dagger$ with
\begin{eqnarray} \label{tran}
S \, b_{\lambda}^{(i) \dagger} \, S^\dagger = \sum_{j=1}^N U_{ij} \, c_{\lambda}^{(i) \dagger} 
\end{eqnarray}
and the leakage operator $L^{(i)}$ becomes
\begin{equation} \label{reset2}
\tilde L^{(i)} =  \sum_{\lambda=+,-} \sum_{j=1}^N U_{ij} \, a^{(i)}_{\lambda} \, c_{\lambda}^{(j) \dagger} ~.
\end{equation} 
Applied to the state of the total system, this operator yields the state of the $N$ atom-cavity systems after the leakage of one photon through the 
outcoupling mirror of cavity $i$.

\section{Initialisation} \label{Initialisation}
 
Let us now describe how the atom-cavity system can be prepared in a highly entangled state $|\phi_{\rm ent}^{\rm ini} \rangle$ which can then be used to 
produce a highly entangled mulitphoton state of the same form at any later time. To achieve this we use only interference effects and postselective measurements\cite{LimPRL,Lim}. Initially, all atoms  should be prepared in the ground state $|v \rangle$ with no photons in the cavity and we denote the corresponding state of subsystem 
$i$ by $|v,0 \rangle_i$. Afterwards, a counterintuitive pulse sequence like the one described in detail in Section \ref{demand}, should be 
applied. This sequence requires a laser pulse with polarisation $\pi$ coupling to the $|v \rangle-|f \rangle$ transition of each atom while the cavity 
couples to the $|f \rangle-|u_\pm \rangle$ transitions and populates the cavity field states with one ``$+$'' and one ``$-$'' polarised photon 
simultaneously.  The corresponding time evolution of atom-cavity system $i$ is governed by the Hamiltonian 
\begin{eqnarray}
H_{\rm ini}^{(i)} &=& {\textstyle {1 \over 2}}  \hbar \Omega^{(i)}_{v}(t) \, |v \rangle_{ii} \langle f| 
+ \sum_{\lambda=+,-}  \hbar g^{(i)}_{u_{\lambda}} \, a_{\lambda}^{(i) \dagger} \, |u_{\lambda} \rangle_{ii} \langle f| 
+ {\rm H.c.} 
\end{eqnarray}   
For simplicity, we assume as in Section 2 that the cavity coupling constants $g^{(i)}_{u_{\lambda}}$ are for all atoms and transitions the same. Using 
Eq.~(\ref{U2}), one can then show that the STIRAP process transfers each atom-cavity system to a very good approximation into an entangled state between 
the atom and its cavity field and the state of the system becomes
\begin{equation} \label{ski}
|\phi_0 \rangle = \prod_{i=1}^N \tilde P_{\rm STIRAP}^{(i)} |v,0 \rangle_i
= {\textstyle {1 \over \sqrt{2^N}}} \, \prod_{i=1}^N \big[ \, |u_+,1_+\rangle_i + |u_-,1_-\rangle_i \, \big] ~.
\end{equation} 
After a certain time, given by the cavity life time $1/\kappa$, all photons leaked out through the outcoupling  mirrors of the resonators into the input ports of the multiport beamsplitter. From there they are distributed to the possible output ports where they cause clicks at polarisation sensitive detectors (see Figure \ref{setup}(b)). 

Here we are interested in the case where exactly one photon arrives at each output port. If this is not the case, the initialisation step should be 
repeated. Proceeding like this has the advantage that the proposed scheme does not require detectors which are able to 
distinguish between states with one, two or more photons. The initialisation step is successful if all $N$ polarisation sensitive photon detectors have 
registered a photon within a small time interval of $1/\kappa$. This also allows us to filter out incidences of dark counts on the detectors and allows us to prepare the initialised entangled atomic state with a high fidelity.

Using Eq. (\ref{reset2}), we can now easily calculate the effect of the detection of a photon at output port $j$  and with polarisation $\lambda_j~(\lambda_j=+,-)$. In the case where one photon is detected at each output port of the multiport beamsplitter via absorption, the state of the system becomes
\begin{equation}
|\phi_{\rm ent}^{\rm ini} \rangle 
= \prod_{j=1}^N  \prod_{i=1}^N \, c^{(j)}_{\lambda_j} \, \tilde L^{(i)} \, |\phi_0 \rangle /\| \cdot \| ~.
\end{equation} 
Note that the final state does not depend on the order in which the detectors register the incoming photons and therefore also not on the order in which 
the leakage operators $\tilde L^{(i)}$ are applied to $|\phi_0 \rangle$. By choosing a suitable linear optics network with appropriate transition matrix 
elements $U_{ij}$ a great variety of highly entangled states between the $N$ atoms in the distant cavities can be produced. For example, the preparation 
of the atoms in a so-called {\em W}-state succeeds in case of the detection of one ``$+$'' polarised photon in one output port and a ``$-$'' polarised 
photon in every other\cite{Lim}. If the initialisation step failed, it has to be repeated until the requested state has been prepared. 

\subsection{Preparing a maximally entangled state of two atoms} 

\begin{figure}
\begin{center}
\begin{tabular}{c}
\includegraphics[height=5cm]{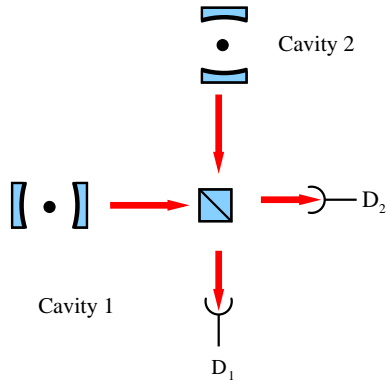}
\end{tabular}
\end{center}
\caption[example]{ \label{exsetup} 
Experimental setup for the generation of a maximally entangled state between two atoms trapped in distant optical cavities. This step requires only a 50:50 beamsplitter and the detectors $D_1$ and $D_2$.}
\end{figure} 

As an example, we now describe the preparation of two atom-cavity systems in a maximally entangled state. This state can then be used as an input state 
for the generation of a maximally entangled photon pair on demand. The corresponding experimental setup for the initialisation step is shown in Figure 
\ref{exsetup}. For the simple case $N=2$, the multiport beamsplitter consists only of a single 50:50 beamsplitter and the transition matrix elements 
$U_{ij}$ become $U_{11}=U_{12}=U_{21}=-U_{22}=1/\sqrt{2}$. 

Initially, the system should be prepared in the state $|v,0 \rangle_1 |v,0 \rangle_2$. Afterwards, the STIRAP process described above, transfers the system 
according to Eq.~(\ref{ski}) into the state
\begin{equation} 
|\phi_0 \rangle = {\textstyle {1 \over 2}} \, \prod_{i=1}^2 \big[ \, |u_+,1_+\rangle_i + |u_-,1_-\rangle_i \, \big] ~.
\end{equation} 
Suppose, a ``$+$'' polarised photon is detected at detector $D_1$ and a ``$-$'' polarised photon is detected in detector $D_2$, then the state of the 
system becomes
\begin{equation} \label{fun}
|\phi_{\rm ent}^{\rm ini} \rangle 
= {\textstyle {1 \over \sqrt{2}}} \, \big[ \, |u_+,0\rangle_1 |u_-,0\rangle_2 - |u_-,0\rangle_1 |u_+,0\rangle_2 \, \big] ~,
\end{equation} 
which describes two atoms in a maximally entangled state. The same maximally entangled state of the atoms is also prepared in the case where a ``$-$'' polarised photon is measured at detector $D_1$ and a ``$+$'' polarised photon is measured at detector $D_2$. The probability that both detectors find a 
photon with the same polarisation equals zero and the probability for the initialisation step to be successful is close to $1/2$ for a wide range of 
experimental parameters.

\section{Mapping} \label{Mapping}

Suppose the system has already been prepared in a highly entangled state $|\psi^{\rm ini}_{\rm ent}\rangle$ involving only the atomic ground states $|u_+ \rangle$ and $|u_- \rangle$ while there are no photons in the resonators. Mapping of this state onto the state of $N$ newly generated photons then requires only a single $\pi$-polarised laser pulse coupling 
simultaneously to the $|u_+\rangle-|e_- \rangle$ and the $|u_- \rangle-|e_+ \rangle$ transition of each atom. At the same time, the cavity field 
should interact with the $|e_+ \rangle-|v \rangle$ and the $|e_- \rangle-|v \rangle$ transition of each atom. The corresponding time evolution for 
atom-cavity system $i$ is then described by the Hamiltonian 
\begin{eqnarray} \label{map H}
H_{\rm map}^{(i)} &=&  {\textstyle {1 \over 2}} \hbar \Omega^{(i)}_{u_+} \, |u_+ \rangle_{ii} \langle e_-| + {\textstyle {1 \over 2}} \hbar 
\Omega^{(i)}_{u_-} \, |u_- \rangle_{ii} \langle e_+| + \hbar g^{(i)}_{e_+} \, a_+^{(i) \dagger} \, |v \rangle_{ii} \langle e_+| + \hbar g^{(i)}_{e_-} \, 
a_-^{(i) \dagger} \, |v \rangle_{ii} \langle e_-| + {\rm H.c.} 
\end{eqnarray} 
The mapping process consists of a simple STIRAP process, as described in Section 2, and should be applied to all atom-cavity systems simultaneously. For 
simplicity, we assume again that the cavity coupling constants $g^{(i)}_{e_{\lambda}}$ are for all transitions and subsystems the same. Then the effect of 
the so-called counterintuitive pulse sequence on system $i$ corresponds in direct analogy to Eq. (\ref{U}) to the operator
\begin{equation}
P_{\rm STIRAP}^{(i)}=|v,1_-\rangle_{ii} \langle u_+,0|+ |v,1_+\rangle_{ii} \langle u_-,0| ~.
\end{equation}
Note that the mapping step transfers each initial state of the form $|u_{\lambda_1},0\rangle_1 |u_{\lambda_2},0 \rangle_2 \, ... \, |u_{\lambda_N},0 \rangle_N$ 
into a state with all atoms in $|v \rangle$ and one photon with $\lambda_1$ in cavity 1, one photon with $\lambda_2$ in cavity 2 and so on. Since the underlying 
STIRAP process is coherent, the final state is a superposition of all these states. 

Once created, one photon leaks out through the outcoupling mirror of each cavity into a separate mode of the free radiation field. Thereby the state of 
the system becomes  
\begin{equation} \label{map}
|\psi^{\rm map}_{\rm ent} \rangle = \prod_{i=1}^N  L^{(i)} \, P_{\rm STIRAP}^{(i)} |\psi^{\rm ini}_{\rm ent}\rangle 
\end{equation}
with $L^{(i)}$ defined by Eq.~(\ref{U}). The probability for the generation of the final multiphoton state (\ref{map}) can, at least in principle, be arbitrarily close to one. 

\subsection{Entangled photon pair creation on demand} 

Suppose no measurements are performed beyond the outcoupling mirrors of the cavities, whether photons have been created or not. If the system has 
initially been prepared in the maximally entangled state (\ref{fun}), the final state (\ref{map}) equals 
\begin{eqnarray}
|\psi^{\rm map}_{\rm ent} \rangle &=& {\textstyle {1 \over \sqrt{2}}}  \, \prod_{i=1}^2 L^{(i)} \, P_{\rm STIRAP}^{(i)} \big[ \, |u_+,0\rangle_1 
|u_-,0\rangle_2 - |u_-,0\rangle_1 |u_+,0\rangle_2 \, \big] \nonumber \\
&=&  {\textstyle {1 \over \sqrt{2}}} \, 
|v,0 \rangle_1 |v,0 \rangle_2 \otimes \big[ \, |1_+\rangle_1 |1_-\rangle_2 - |1_-\rangle_1 |1_+\rangle_2 \, \big] ~,
\end{eqnarray}
which describes indeed a maximally entangled photon pair in the free radiation field. The fidelity of the generated entanglement is high unless the atoms 
suffer from decoherence before the mapping stage is performed. However, this decoherence time can be very long since the initial state (\ref{fun}) is an atomic ground state and its life time is not restricted by spontaneous emission from the atoms.

\section{Conclusion}

In recent years, several proposals for the measurement-induced generation of entanglement have been made.\cite{Cabrillo,Cabrillo2,Cabrillo3,Cabrillo4,Cabrillo5} Other authors analysed the possibility to map the state of an atom-cavity system onto a photon\cite{Parkins,Parkins2}. Here we combine all these ideas with a scheme for the generation of single photons on demand\cite{Law,Kuhn} and propose an experiment for the generation of multiphoton entanglement on demand.  A more detailed description of the presented scheme will be published elsewhere \cite{Lim}.

\acknowledgments
Y. L. L. thanks the DSO National Laboratories in Singapore for funding as a PhD student. A. B. acknowledges a James Ellis University Research Fellowship 
from the Royal Society and the GCHQ. This work was supported in part by the European Union and the UK Engineering and Physical Sciences Research Council.

\end{document}